# Semi-quantum dialogue based on single photons

Tian-Yu Ye*, Chong-Qiang Ye

College of Information & Electronic Engineering, Zhejiang Gongshang University, Hangzhou 310018, P.R.China

**Abstract:** In this paper, we propose two semi-quantum dialogue (SQD) protocols by using single photons as the quantum carriers, where one requires the classical party to possess the measurement capability and the other does not have this requirement. The security towards active attacks from an outside Eve in the first SQD protocol is guaranteed by the complete robustness of present semi-quantum key distribution (SQKD) protocols, the classical one-time pad encryption, the classical party's randomization operation and the decoy photon technology. The information leakage problem of the first SQD protocol is overcome by the classical party' classical basis measurements on the single photons carrying messages which makes him share their initial states with the quantum party. The security towards active attacks from Eve in the second SQD protocol is guaranteed by the classical party's randomization operation, the complete robustness of present SQKD protocol and the classical one-time pad encryption. The information leakage problem of the second SQD protocol is overcome by the quantum party' classical basis measurements on each two adjacent single photons carrying messages which makes her share their initial states with the classical party. Compared with the traditional information leakage resistant QD protocols, the advantage of the proposed SQD protocols lies in that they only require one party to have quantum capabilities. Compared with the existing SQD protocol, the advantage of the proposed SQD protocols lies in that they only employ single photons rather than two-photon entangled states as the quantum carriers. The proposed SQD protocols can be implemented with present quantum technologies.

## 1 Introduction

Quantum cryptography, derived from Bennett and Brassard's pioneering work on quantum key distribution (QKD) protocol [1] in 1984, utilizes the property of quantum mechanics rather than the computation difficulty of mathematical problems to attain the unconditional security. It has attracted lots of attention and established many interesting branches, such as QKD [1-5], quantum secure direct communication (QSDC) [6-13], quantum secret sharing (QSS) [14-18] *etc*.

QKD is devoted to establishing a sequence of random key between two remote parties through the transmission of quantum signals while QSDC concentrates on transmitting a confidential message directly from one party to the other one without establishing a sequence of random key first. In 2004, in order to accomplish the mutual exchange of secret messages from two parties, Zhang and Man [19-20] and Nguyen [21] independently suggested the novel concept named quantum dialogue (QD). As a result, QD has greatly aroused the interest of researchers. However, the earliest QD protocols [19-27] always have the problem named information leakage, which means that anyone else can easily extract some useful information about the secret messages of two parties without launching any active attacks. The information leakage problem existing in QD was independently discovered by Gao *et al.* [28] and Tan and Cai [29] in 2008. Subsequently, researchers quickly turned to study how to solve the information leakage problem existing in QD. Until now, many excellent methods have been suggested, such as utilizing the direct transmission of auxiliary quantum states [30-37], the correlation extractability of Bell states[38], controlled-not operations and auxiliary single photons [39], the measurement correlation from entanglement swapping of quantum entangled states[40-41], the encoding for collections composed by the entanglement swapping results of quantum entangled states[34-36], quantum encryption sharing[42-43], auxiliary quantum operations[44] and the measurement correlation of quantum entangled states[45].

In 2007, by using the famous BB84 protocol [1], Boyer *et al.* [46] put forward the first semi-quantum cryptography protocol (i.e., the BKM2007 protocol) which allows that only one party possesses quantum capabilities. In the BKM2007 protocol, the receiver Bob is restricted to perform the following operations in quantum channel: (a) sending or returning the qubits without disturbance; (b) measuring the qubits in the fixed orthogonal basis $\{|0\rangle,|1\rangle\}$; (c) preparing the (fresh) qubits in the fixed orthogonal basis $\{|0\rangle,|1\rangle\}$. In 2009, Boyer *et al.* [47] constructed a semi-quantum key distribution (SQKD) protocol based on randomization by using single photons, where the receiver Bob is restricted to perform (a), (b) and (d) reordering the qubits (via different delay lines). According to the definition of the protocols in Refs.[46-47], the orthogonal basis $\{|0\rangle,|1\rangle\}$ can be regarded as a classical basis as it only refers to qubits $|0\rangle$ and $|1\rangle$ rather than any quantum superposition state, and can be replaced with the classical notation $\{0,1\}$. Moreover, the receiver Bob is called classical as he is restricted to perform the above four operations, namely, (a), (b), (c) and (d). Apparently, different from the traditional quantum cryptography which requires all parties to possess quantum capabilities, semi-quantum cryptography allows partial parties to merely possess classical capabilities rather than quantum capabilities so that they needn't refer to the preparation and measurement of quantum superposition states. Therefore, semi-quantum cryptography is favorable for partial parties to ease the burdens of quantum state preparation and measurement.

Since the novel concept of "semi-quantumness" was first proposed by Boyer *et al.* [46] in 2007, researchers have shown great enthusiasm on it and applied it onto different quantum cryptography tasks such as QKD, QSDC and QSS. As a result, many

---

*Corresponding author:

　E-mail：happyyty@aliyun.com(T.Y.Ye)

semi-quantum cryptography protocols, such as SQKD protocols [46-63], semi-quantum secure direct communication (SQSDC) protocols [50,64], semi-quantum secret sharing (SQSS) protocols [65-69], semi-quantum private comparison (SQPC) protocols [70-71], semi-quantum key agreement (SQKA) protocols [72-73], controlled deterministic secure semi-quantum communication (CDSSQC) [73] protocol, semi-quantum dialogue (SQD) protocol [73] *etc*, have been suggested.

It is easy to discover that all of the above QD protocols [19-27,30-45] require both parties to possess quantum capabilities which may be unpractical in some circumstances, as not both parties have the abilities to afford expensive quantum resources and operations. Whether the dialogue can be successfully accomplished if there is only one party who has quantum capabilities? The first SQD protocol suggested in Ref.[73] by using Bell entangled states gave this question a positive answer.

Based on the above analysis, in this paper, we are devoted to designing two SQD protocols by using single photons as the quantum carriers, where one requires the classical party to possess the measurement capability and the other does not have this requirement. Compared with the traditional information leakage resistant QD protocols, the advantage of the proposed SQD protocols lies in that they only require one party to have quantum capabilities. Compared with the existing SQD protocol, the advantage of the proposed SQD protocols lies in that they only employ single photons rather than two-photon entangled states as the quantum carriers.

The rest of this paper is arranged as follows: In Section 2, two SQD protocols with single photons requiring and without requiring the measurement capability of classical party are designed, respectively, and their security is analyzed; in Section 3, discussion and conclusion are given.

## 2 The SQD protocols based on single photons

Suppose that there are two parties, Alice and Bob, each of whom has $N$ bits secret messages. Alice has quantum capabilities while Bob is restricted to only possess classical capabilities. They want to exchange their secret messages by using single photons as the quantum carriers. They agree on beforehand that for encoding, the unitary operation $I$ ($\sigma_x$) stands for the classical bit 0 (1), where $I = |0\rangle\langle 0| + |1\rangle\langle 1|$ and $\sigma_x = |0\rangle\langle 1| + |1\rangle\langle 0|$; and for decoding, the qubit $|0\rangle$ ($|1\rangle$) stands for the classical bit 0 (1).

In this section, in order to accomplish the task of exchanging Alice and Bob's secret messages, we construct two SQD protocols by using single photons. The first one requires the classical party to possess the measurement capability while the second one does not have this requirement.

### 2.1 The SQD protocol based on single photons requiring the measurement capability of classical party
### A. Protocol description

In the SQSDC protocol of Ref.[64], the classical party can successfully transmit his secret messages to the quantum party, where the classical party needs to measure the qubits sent from the quantum party. In order to accomplish the bidirectional communication, inspired by this protocol, we suggest a SQD protocol based on single photons requiring the measurement capability of classical party which is composed by the following steps.

**Step 1:** Quantum Alice prepares $8N$ single photons, each of which is randomly in one of the four states $\{|0\rangle, |1\rangle, |+\rangle, |-\rangle\}$, and transmits them to classical Bob in the way of block transmission [6].

**Step 2:** After receiving all single photons from Alice, Bob implements the following security check procedures together with Alice: (1) Bob randomly picks out half of the $8N$ single photons; (2) As for each chosen photon, Bob randomly reflects it to Alice, or measures it with $Z$ basis and resends it to Alice in the same state he found; (3) After getting Alice's announcement of receipt, Bob tells Alice the positions where he chose to reflect and where he chose to measure as well as the measurement results; (4) For the reflected photons, Alice measures them with her preparation basis and calculates the error rate by comparing the measurement results with her prepared states; and for the photons both prepared by her in $Z$ basis and measured by Bob, Alice measures the corresponding returned photons with $Z$ basis and calculates the corresponding error rate; (5) If the total error rate is lower than the threshold, they will continue the communication and proceed to the next step; otherwise, they will terminate the communication.

**Step 3:** For the remaining $4N$ single photons, Alice publishes which ones were prepared by her in $Z$ basis. It is expected that there are approximately $2N$ single photons in $Z$ basis on Bob's site now.

**Step 4:** From the single photons in $Z$ basis on his site, Bob randomly picks out $N$ ones to encode his secret messages. These chosen single photons are called as single photons carrying messages. The left single photons on Bob's site are called as sample single photons, which will be used for security check below. Bob's encoding rule is: if his secret classical bit is 0, Bob will measure the single photon carrying messages with $Z$ basis and prepare a fresh one in the same state; and if his secret classical bit is 1, Bob will measure the single photon carrying messages with $Z$ basis and prepare a fresh one in the opposite state. Note that Bob can know the initial states of single photons carrying messages with his $Z$ basis measurements. Finally, Bob randomly reorders all single photons in his hand and sends them to Alice in the way of block transmission [6].

**Step 5:** After getting Alice's announcement of receipt, Bob publishes the orders of the transmitted single photons and the positions of single photons carrying messages. Alice restores all single photons in the original orders according to Bob's announcement. For the sample single photons, it is expected that there are approximately $N$ ones in $Z$ basis and $2N$ ones in $X$ basis (i.e., the orthogonal basis $\{|+\rangle, |-\rangle\}$). Alice implements the security check procedure with these sample single photons, i.e., she measures them with her preparation basis and calculates the error rate by comparing the measurement results with her prepared states. If the error rate is lower than the threshold, she will continue the communication and proceed to the next step; otherwise, she will terminate the communication.

**Step 6:** Alice discards all sample single photons. Then Alice encodes her secret messages on the single photons carrying messages according to the following rule: If her secret classical bit is 0, Alice will keep the single photon carrying messages unchanged; and if her secret classical bit is 1, Alice will perform the unitary operation $\sigma_x$ on the single photon carrying messages. Afterward, Alice measures all single photons carrying messages with $Z$ basis and publishes her measurement results on their final states to Bob. According to his own secret messages, the final states of single photons carrying messages and his knowledge on their initial states, Bob can read out Alice's secret messages. In the meanwhile, according to her own secret messages, the final states of single photons carrying messages and their prepared initial states, Alice can read out Bob's secret messages.

For clarity, we depict the flow chart of the above SQD protocol in Fig.1.

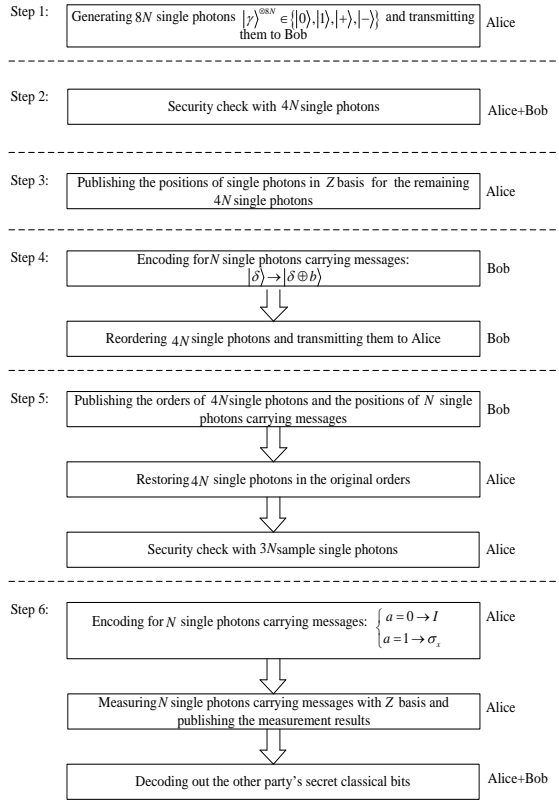

Fig. 1 The flow chart of the proposed SQD protocol

"$|\delta\rangle \rightarrow |\delta \oplus b\rangle$" means that Bob measures $|\delta\rangle$ with $Z$ basis and prepares a fresh one in the state of $|\delta \oplus b\rangle$, where $b$ is Bob's secret classical bit, $|\delta\rangle$ is the single photon carrying messages, $\oplus$ is the XOR operation and $\delta, b \in \{0,1\}$. Moreover, "$\begin{cases} a=0 \rightarrow I \\ a=1 \rightarrow \sigma_x \end{cases}$" means that if $a=0$, Alice will keep the single photon carrying messages unchanged; and if $a=1$, Alice will perform the unitary operation $\sigma_x$ on the single photon carrying messages, where $a$ is Alice's secret classical bit.

In the above protocol, Alice needs to prepare and measure the qubits in $X$ basis. As a result, Alice is required to have quantum capabilities. On the other hand, Bob is restricted to perform the following four operations: 1) measuring the qubits in $Z$ basis; 2) preparing the (fresh) qubits in $Z$ basis; 3) reordering the qubits (via different delay lines); 4) sending or returning the qubits without disturbance. As a result, Bob is not necessarily required to have quantum capabilities. Therefore, the above protocol is indeed a SQD protocol requiring the measurement capability of classical party.

For clarity, we summarize the encoding and decoding rules of the above SQD protocol in Table 1. We use an example to further illustrate its bidirectional communication process without considering the security check procedures. Suppose that Bob has four secret classical bits '0011' and Alice has four secret classical bits '0101'. Moreover, assume that the four single photons carrying messages prepared by Alice in Step 1 are $\{|0\rangle,|1\rangle,|0\rangle,|1\rangle\}$. In Step 4, Bob can know the initial states of these four single photons carrying messages with his $Z$ basis measurements. After Bob's encoding in this step, the four single photons carrying messages are changed into $\{|0\rangle,|1\rangle,|1\rangle,|0\rangle\}$. Further, after Alice's encoding in Step 6, the four single photons carrying messages are turned into $\{|0\rangle,|0\rangle,|1\rangle,|1\rangle\}$. Alice measures these four single photons carrying messages with $Z$ basis and publishes their final states to Bob. According to his own secret classical bits, the final states of these four single photons carrying messages and his knowledge on their initial states, Bob can read out that Alice's secret classical bits are '0101'. In the meanwhile, according to her own secret classical bits, the final states of these four single photons carrying messages and her prepared initial states on them, Alice can read out that Bob's secret classical bits are '0011'. It can be concluded that Alice and Bob have successfully exchanged their secret classical bits.

**B.  Security analysis**

(1) Analysis on the information leakage problem

We analyze the security towards a passive attack named information leakage here. In Step 4, Bob can know the initial states of single photons carrying messages with his $Z$ basis measurements. As a result, it is not necessary for Alice to publish their initial states. Consequently, Eve has no chance to know their initial states from Alice or Bob's public announcement. From the viewpoint of Shannon's information theory [74], for each single photon carrying messages, Alice's public announcement on its final state totally involves four kinds of possibilities on Alice and Bob's secret classical bits, containing $-\sum_{i=1}^{4} p_i \log_2 p_i = -4 \times \frac{1}{4} \log_2 \frac{1}{4} = 2$ bit information for Eve. Therefore, no information leaks out to Eve. Obviously, Bob's $Z$ basis measurements on the single photons carrying messages in Step 4 make him share their initial states with Alice, which helps overcome the information leakage problem.

Table 1    The encoding and decoding table

| Single photon carrying messages prepared by Alice | Bob's secret classical bit | Single photon carrying messages after Bob's encoding | Alice's secret classical bit | Single photon carrying messages after Alice's encoding |
|---|---|---|---|---|
| $|0\rangle$ | 0 | $|0\rangle$ | 0 | $|0\rangle$ |
|  |  |  | 1 | $|1\rangle$ |
|  | 1 | $|1\rangle$ | 0 | $|1\rangle$ |
|  |  |  | 1 | $|0\rangle$ |
| $|1\rangle$ | 0 | $|1\rangle$ | 0 | $|1\rangle$ |
|  |  |  | 1 | $|0\rangle$ |
|  | 1 | $|0\rangle$ | 0 | $|0\rangle$ |
|  |  |  | 1 | $|1\rangle$ |

(2) Analysis on the active attacks from an outside Eve

We analyze the security towards Eve's active attacks here.

Alice transmits $8N$ single photons to Bob in Step 1 and implements the security check procedures together with Bob in Step 2. It is obvious that the security check method of Step 2 is similar to that of SQKD protocols in Refs.[46-47] which has been proven to be completely robust. The difference between the security check method of Step 2 and that of SQKD protocols in Refs.[46-47] is that partial single photons are stored by the receiver in the former, whereas all single photons are returned to the sender in the latter. It is straightforward that the remaining $4N$ single photons in Step 3 are completely robust, which makes Eve have no knowledge about the initial states of single photons carrying messages even if she escapes from the security check procedures. In Step 4, Bob encodes his secret messages on single photons carrying messages. Apparently, this process is identical to the classical one-time pad encryption where Bob's secret messages are encrypted with the initial states of single photons carrying messages. Thus, Eve cannot obtain Bob's secret messages even though she intercepts the encoded single photons carrying messages, because she has no idea about their initial states which are completely random in $Z$ basis.

In addition, as the remaining $4N$ single photons in Step 3 are completely robust, the security check method of Step 5 is identical to the famous decoy photon technology [75-76], which can be seen as a variant of the BB84 eavesdropping check method [1] whose unconditional security has been validated [77]. Consequently, after Bob randomly reorders all single photons in his hand in Step 4, which makes Eve unaware about the positions of single photons carrying messages and sample single photons before Bob publishes them in Step 5, Eve will inevitably leave her trace on the sample single photons and be detected by this security check method if she launches active attacks during the transmission of these single photons.

It can be concluded now that Eve' active attacks are invalid to the above protocol, and that its security towards Eve' active attacks is guaranteed by the complete robustness of present SQKD protocols, the classical one-time pad encryption, the classical party's randomization operation and the decoy photon technology.

### C. The information-theoretical efficiency

The information-theoretical efficiency [4] is defined as $\eta = \frac{b_s}{q_t + b_t} \times 100\%$, where $b_s$, $q_t$ and $b_t$ are the expected secret bits received, the qubits used and the classical bits exchanged between Alice and Bob. In the above protocol, without taking the security check processes into account, $N$ single photons carrying messages can be used for encoding $2N$ secret classical bits from Alice and Bob. In the meanwhile, Bob needs to prepare $N$ fresh single photons carrying messages when encoding his $N$ secret classical bits. In addition, $N$ classical bits are needed for Alice to inform Bob of her measurement results on the final states of $N$ single photons carrying messages. Hence, the information-theoretical efficiency of the above protocol is $\eta = \frac{2N}{N+N+N} \times 100\% \approx 66.7\%$.

## 2.2 The SQD protocol based on single photons without requiring the measurement capability of classical party
### A. Protocol description

The SQKD protocols in Refs.[50-52] all release the classical party from the classical basis measurement. Inspired by these protocols, we suggest a SQD protocol based on single photons also without requiring the measurement capability of classical party which is composed by the following steps.

**Step 1:** Quantum Alice prepares $N$ single photons, each of which is randomly in one of the four states $\{|0\rangle,|1\rangle,|+\rangle,|-\rangle\}$, and transmits them to classical Bob in the way of block transmission [6].

**Step 2:** For encoding his own secret messages, Bob prepares $2N$ single photons carrying messages in $Z$ basis: if his secret classical bit is 0, he will prepare two adjacent single photons carrying messages in the same states; and if his secret classical bit is 1, he will prepare two adjacent single photons carrying messages in the opposite states. Moreover, Bob prepares $M+2N$ ($M \geq N$) sample single photons randomly in $Z$ basis. Bob selects $2N$ sample single photons and uses them to compose a qubit string $|\varphi\rangle$ with $2N$ single photons carrying messages. After receiving $N$ single photons from Alice, Bob uses them to compose a qubit string $|\phi\rangle$ with the remaining $M$ sample single photons. Then, Bob randomly selects the first $2N$ single photons $|\psi\rangle$ from qubit string $|\phi\rangle$. Finally, Bob reorders the single photons in $|\varphi\rangle$ and $|\psi\rangle$, and sends them to Alice in the way of block transmission [6]. For convenience, in $|\psi\rangle$, we call the single photons from Alice *CTRL single photons* and the single photons prepared by Bob *SIFT single photons*.

**Step 3:** Bob publishes the positions of single photons which belong to $|\psi\rangle$. Alice measures each single photon in $|\psi\rangle$ randomly with $Z$ basis or $X$ basis.

**Step 4:** Bob publishes the orders of single photons in $|\psi\rangle$. In the meanwhile, Alice publishes the positions in $|\psi\rangle$ where she measured with $Z$ basis. $Z$-SIFT bits denote the bits produced by Alice using $Z$ basis to measure the STFT single photons.

**Step 5:** Alice checks the error rate on the CTRL single photons in $|\psi\rangle$ by comparing her measurement results with her prepared states. If Alice selected the correct basis to measure the CTRL single photons in $|\psi\rangle$ in Step 3, her measurement results should be the same as her prepared states. If the error rate is higher than some predefined threshold $P_{CTRL}$, they will terminate the protocol.

**Step 6:** Alice publishes the values of $Z$-SIFT bits in $|\psi\rangle$. Alice's measurement results should be the same as Bob's prepared states. Bob checks the error rate on $Z$-SIFT bits in $|\psi\rangle$. If the error rate is higher than some predefined threshold $P_{Z-SIFT}$, they will terminate the protocol.

**Step 7:** Alice discards the single photons which belong to $|\psi\rangle$. Bob publishes the positions and the orders of $2N$ single photons carrying messages in $|\varphi\rangle$. After Bob's announcement on them, Alice discards the $2N$ sample single photons in $|\varphi\rangle$ and reorders the remaining $2N$ single photons carrying messages in the original orders. Then Alice measures each two adjacent single photons carrying messages and reads out Bob's secret messages. Concretely, if each two adjacent single photons carrying messages measured by her are in the same states, Alice will know that Bob's secret classical bit is 0; and if each two adjacent single photons carrying messages measured by her are in the opposite states, Alice will know that Bob's secret classical bit is 1. For transmitting her secret messages to Bob, Alice implements the classical one-time pad encryption processes and publishes the ciphertexts to Bob according to the following rules: if the second photon of each two adjacent single photons carrying messages she measured is in the state of $|0\rangle$ and her secret classical bit is 0, she will publish 0 to Bob; and if the second photon of each two adjacent single photons carrying messages she measured is in the state of $|0\rangle$ and her secret classical bit is 1, she will publish 1 to Bob; and if the second photon of each two adjacent single photons carrying messages she measured is in the state of $|1\rangle$ and her secret classical bit is 0, she will publish 1 to Bob; and if the second photon of each two adjacent single photons carrying messages she measured is in the state of $|1\rangle$ and her secret classical bit is 1, she will publish 0 to Bob. According to the ciphertexts announced by Alice and the classical bits represented by all of the second photons from each two adjacent single photons carrying messages prepared by himself, Bob can directly decode out Alice's secret messages.

For clarity, we depict the flow chart of the above SQD protocol in Fig.2.

In the above protocol, Alice needs to prepare and measure the qubits in $X$ basis. As a result, Alice is required to have quantum capabilities. On the other hand, Bob is restricted to perform the following three operations: 1) preparing the qubits in $Z$ basis; 2) reordering the qubits (via different delay lines); 3) sending or returning the qubits without disturbance. As a result, Bob is required to have neither quantum capabilities nor the measurement capability. Therefore, the above protocol is indeed a SQD protocol without requiring the measurement capability of classical party.

For clarity, we summarize the encoding and decoding rules of the above SQD protocol in Table 2. We use an example to further illustrate its bidirectional communication process without considering the security check procedures. Suppose that Bob has four secret classical bits '0011' and Alice has four secret classical bits '0101'. Moreover, assume that according to his own secret classical bits, Bob prepares eight single photons carrying messages in the states of $\{|0\rangle,|0\rangle,|1\rangle,|1\rangle,|0\rangle,|1\rangle,|1\rangle,|0\rangle\}$ in Step 2. After Alice measures each two adjacent single photons carrying messages in Step 7, she can directly read out that Bob's secret classical bits are '0011'. After Alice publishes the ciphertexts '0011' to Bob in Step 7, Bob can directly know that Alice's secret classical bits are '0101'.

**B. Security analysis**

(1) Analysis on the information leakage problem

We analyze the security towards information leakage here. In Step 7, Alice measures each two adjacent single photons carrying messages and reads out Bob's secret messages. As a result, it is not necessary for Bob to publish the initial states of each two adjacent single photons carrying messages. Consequently, Eve has no chance to know the initial states of each two adjacent single photons carrying messages from Alice or Bob's public announcement. From the viewpoint of Shannon's information theory [74], for each two adjacent single photons carrying messages, Alice's public announcement of the ciphertext totally involves four kinds of possibilities on Alice and Bob's secret classical bits, containing $-\sum_{i=1}^{4} p_i \log_2 p_i = -4 \times \frac{1}{4} \log_2 \frac{1}{4} = 2$ bit information for Eve. Therefore, no information leaks out to Eve. Obviously, Alice's measuring each two adjacent single photons carrying messages makes her share their initial states with Bob, which helps overcome the information leakage problem.

(2) Analysis on the active attacks from an outside Eve

We analyze the security towards Eve's active attacks here.

In the above SQD protocol, the single photons in $|\psi\rangle$ are used for security check. Apparently, this security check method is similar to that of Ref.[51] which has been proven to be completely robust. The difference between this security check method and that of Ref.[51] is that partial single photons sent from the receiver are not measured by the sender in the former, whereas all single photon sent from the receiver are measured by the sender in the latter. Before Bob publishes the positions of single photons which belong to $|\psi\rangle$ in Step 3, among the single photons sent from Bob to Alice, Eve cannot discriminate which ones belong to $|\psi\rangle$ and which ones belong to $|\varphi\rangle$. Thus, Eve's attack on the single photons sent from Bob to Alice is independent from the positions for $|\psi\rangle$ and $|\varphi\rangle$. It is straightforward that the single photons in $|\varphi\rangle$ are completely robust, which makes Eve have no knowledge about their initial states even if she escapes from the security check procedures. Accordingly, after Bob publishes the positions and the orders of $2N$ single photons carrying messages in $|\varphi\rangle$ in Step 7, Eve cannot obtain Bob's secret messages. On the other hand, Eve may intercept the single photons sent from Bob to Alice in Step 2 and resend a sequence of fake single photons prepared by herself to Alice. However, Eve still cannot know the genuine positions and orders of $2N$ single photons carrying messages after Bob publishes the positions of single photons which belong to $|\psi\rangle$ in Step 3, due to Bob's randomization operation in Step 2. Thus, Eve cannot obtain Bob's secret messages by intercepting and measuring the single photons sent from Bob to Alice. Note that Eve's intercept-resend attack can be easily detected by the security check procedures, since it may alter the states of single photons from Alice.

In addition, in Step 7, Alice encrypts her secret messages with the classical one-time pad processes where the second photons from each two adjacent single photons carrying messages she measured play the role of private key. However, the security check procedures guarantee that Eve is unable to know the initial states of each two adjacent single photons carrying messages prepared by Bob even if she escapes from the security check procedures. Consequently, Eve cannot decrypt out Alice's secret messages.

It can be concluded now that Eve' active attacks are invalid to the above protocol, and that its security towards Eve' active attacks is guaranteed by the classical party's randomization operation, the complete robustness of present SQKD protocol and the classical one-time pad encryption.

**C. The information-theoretical efficiency**

In the above protocol, without taking the security check processes into account, $2N$ single photons carrying messages can be used for encoding $2N$ secret classical bits from Alice and Bob. In addition, $N$ classical bits are needed for Alice to publish the $N$ bits ciphertexts to Bob. Hence, the information-theoretical efficiency of the above protocol is $\eta = \frac{2N}{2N+N} \times 100\% \approx 66.7\%$.

**3 Discussion and Conclusion**

It is apparent that the qubit transmissions are in a round trip in both of the proposed SQD protocols. As a result, the Trojan horse attacks from Eve, i.e., the invisible photon eavesdropping attack [78] and the delay-photon Trojan horse attack [79-80], should be taken into account. The way to resist the invisible photon eavesdropping attack is that the receiver inserts a filter in front of his devices to filter out the photon signal with an illegitimate wavelength before he deals with it [80-81]. The way to resist the delay-photon Trojan horse attack is that the receiver uses a photon number splitter (PNS) to split each sample quantum signal into two pieces and measures the signals after the PNS with proper measuring bases [80-81]. If the multiphoton rate is unreasonably high, this attack will be detected so that the communication will be terminated.

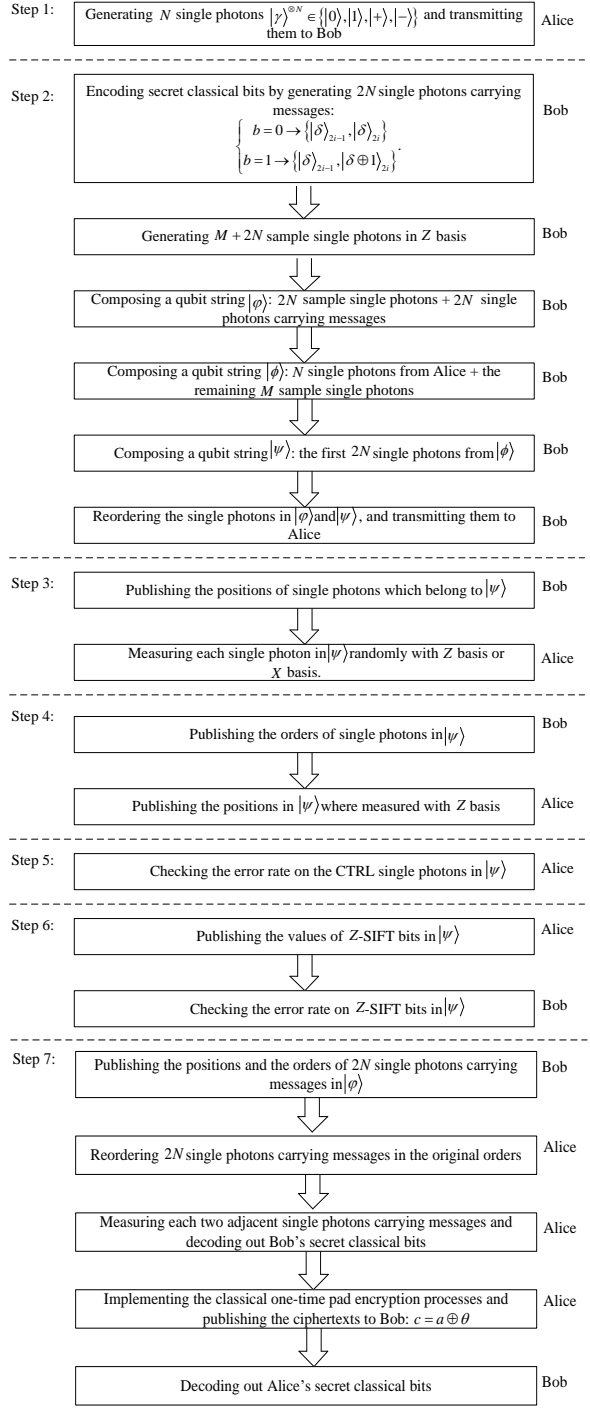

Fig. 2 The flow chart of the proposed SQD protocol

" $\begin{cases} b=0 \to \{|\delta\rangle_{2i-1}, |\delta\rangle_{2i}\} \\ b=1 \to \{|\delta\rangle_{2i-1}, |\delta\oplus 1\rangle_{2i}\} \end{cases}$ " means that if $b=0$, Bob will prepare two adjacent single photons carrying messages in the same states; and if $b=1$, Bob will prepare two adjacent single photons carrying messages in the opposite states. Here, $b$ is Bob's secret classical bit, $\delta \in \{0,1\}$ and $i \in \{1,2,\ldots,N\}$. Moreover, among " $c = a \oplus \theta$ ", $a$ is Alice's secret classical bit, $|\theta\rangle$ is the second photon of each two adjacent single photons carrying messages, $c$ is the ciphertext and $a, \theta, c \in \{0,1\}$.

Table 2  The encoding and decoding table

| Bob's secret classical bit | Two adjacent single photons carrying messages prepared by Bob | | Alice's secret classical bit | The ciphertext published by Alice |
|---|---|---|---|---|
| 0 | $|0\rangle$ | $|0\rangle$ | 0 | 0 |
|   |             |             | 1 | 1 |
|   | $|1\rangle$ | $|1\rangle$ | 0 | 1 |
|   |             |             | 1 | 0 |
| 1 | $|0\rangle$ | $|1\rangle$ | 0 | 1 |
|   |             |             | 1 | 0 |
|   | $|1\rangle$ | $|0\rangle$ | 0 | 0 |
|   |             |             | 1 | 1 |

It is apparent that the SQD protocol of Ref.[73] uses Bell entangled states as the quantum carries while the proposed SQD protocols employ single photons as the quantum carries. As the preparation of single photons is easier than that of Bell entangled states, the proposed SQD protocols exceed the SQD protocol of Ref.[73] in the quantum carriers.

In addition, the proposed SQD protocols need to prepare, measure and store single photons. As a result, they require the quantum technologies for preparing, measuring and storing single photons. Fortunately, these quantum technologies are available at present. Therefore, they have good implementation feasibility currently.

To sum up, by using single photons as the quantum carriers, we propose two SQD protocols, where one requires the classical party to possess the measurement capability and the other does not have this requirement. The proposed SQD protocols can prevent the active attacks from Eve and overcome the information leakage problem. Compared with the traditional information leakage resistant QD protocols, the proposed SQD protocols are more suitable for the circumstance where one party has quantum capabilities and the other one only has classical capabilities. Compared with the existing SQD protocol, the advantage of the proposed SQD protocols lies in that they only employ single photons rather than two-photon entangled states as the quantum carriers. The proposed SQD protocols can be implemented with present quantum technologies.

**Acknowledgments**
   The authors would like to thank the anonymous reviewers for their valuable comments that help enhancing the quality of this paper. Funding by the National Natural Science Foundation of China (Grant No.61402407) and the Natural Science Foundation of Zhejiang Province (Grant No.LY18F020007) is gratefully acknowledged.